\documentclass[twocolumn,showpacs,aps,prl,superscriptaddress,floatfix]{revtex4}

\usepackage{graphicx}
\usepackage{epsfig}
\usepackage{amsmath}
\usepackage{amssymb}
\usepackage{hhline}
\usepackage{multirow}
\usepackage{fancyheadings}
\RequirePackage{xspace}

\input {pubboard/babarsym.tex}
\newcommand{\pmin}{0.6}
\newcommand{\avgTrkEff}{97.1}
\newcommand{\percCascBgHighP}{0.8}  
\newcommand{\percCascBg}{2}  
\newcommand{\reductionOaCut}{24}  
\def\valBrPartial {\ensuremath{(10.24 \pm 0.17\stat \pm 0.26\syst)\%} }
\newcommand{\valEpsBrem}{(2.20 \pm 0.35)\%}
\newcommand{\valNprompt}{25{,}070\pm 410\stat}
\def\valBr {\ensuremath{(10.87 \pm 0.18\stat \pm 0.30\syst)\%}}
\def\valVcb {$0.0423 \pm 0.0007(\text{exp}) \pm 0.0020(\text{theory})$}
\newcommand{\valKappa}{1.061 \pm 0.009}

\newcommand{\eD}{\end{document}}

\newcommand{\BRFU}{\ensuremath{\BR(\B \to X_u  e \nu)}}
\newcommand{\BRFC}{\ensuremath{\BR(\B \to X_c  e \nu)}} 
\newcommand{\BRF}{\ensuremath{\BR(\B \to X  e \nu)}}

\newcommand{\mbtox}{\ensuremath{\B \to X e \nu}}

\newcommand{\ls}{\ensuremath{e^\pm e^\pm\ }}
\newcommand{\uls}{\ensuremath{e^+ e^-\ }}

\def\ps   {\ensuremath{\rm \,ps}\xspace}
\def\BR         {{\ensuremath{\cal B}\xspace}}
\def\B       {\ensuremath{B}\xspace}
\def\Bzb     {\ensuremath{\Bbar^0}\xspace}
\def\Bbar    {\kern 0.18em\overline{\kern -0.18em B}{}\xspace}

\def\epem       {\ensuremath{e^+e^-}\xspace}

\def\jpsi     {\ensuremath{{J\mskip -3mu/\mskip -2mu\psi\mskip 2mu}}\xspace}

\def\Vcb  {\ensuremath{|V_{cb}|}\xspace}
\def\Vub  {\ensuremath{|V_{ub}|}\xspace}

\def\invfb   {\ensuremath{\mbox{\,fb}^{-1}}\xspace}

\begin{document}
\noindent BABAR-PUB-02-011 \\
SLAC-PUB-9306 \\

\title[Short Title]{\large \bf
Measurement of the Branching Fraction
for Inclusive Semileptonic \B Meson Decays
 }

%% author list as of 05-Jul-2002 (556 authors)
%
\author{B.~Aubert}
\author{D.~Boutigny}
\author{J.-M.~Gaillard}
\author{A.~Hicheur}
\author{Y.~Karyotakis}
\author{J.~P.~Lees}
\author{P.~Robbe}
\author{V.~Tisserand}
\author{A.~Zghiche}
\affiliation{Laboratoire de Physique des Particules, F-74941 Annecy-le-Vieux, France }
\author{A.~Palano}
\author{A.~Pompili}
\affiliation{Universit\`a di Bari, Dipartimento di Fisica and INFN, I-70126 Bari, Italy }
% ...
\collaboration{The \babar\ Collaboration}
\noaffiliation

\date{\today}

\begin{abstract}
A largely model-independent measurement of the inclusive electron momentum spectrum 
and branching fraction for semileptonic decays of \B mesons is presented 
based on data recorded at the \FourS\ resonance with the \babar\ detector.
Backgrounds from secondary charm decays are separated from prompt \B decays 
using charge and angular correlations between the electron from one \B meson and
a high momentum electron tag from the second \B meson.
The resulting branching fraction is $\BR(\B \to X e \nu) =$ \valBr.
Based on this measurement we determine the CKM matrix element $|V_{cb}|$. 

\end{abstract}

\pacs{12.15.Hh, 11.30.Er, 13.25.Hw}% PACS, the Physics and Astronomy Classification Scheme.

\maketitle

Measurements of semileptonic \B meson decays are a good way to determine 
the CKM matrix elements \Vcb and \Vub, two of the parameters of the Standard Model. 
For \Vcb, analyses of exclusive and inclusive decays have resulted in comparable precision.  
While most measured values of \BRF\ are below 11\%~\cite{PDG00}, theoretical calculations including perturbative QCD contributions predict values of 12\% or above~\cite{slcalc}.

%-----------------------------------------------------------------------------------------
%-----------------------------------------------------------------------------------------

The measurement presented here employs the method introduced by ARGUS \cite{ARGUS93} and later used
by CLEO \cite{CLEO96}, in which \BB\ events are tagged by the presence of a high momentum lepton. 
As a tag, we choose electrons with momentum $p^*$ in the interval 1.4 to 2.3 \gevc, where $p^*$ is 
measured in the center-of-mass frame.
A second electron in the event is taken as the signal lepton for which we require $p^* > \pmin \gevc$, to avoid
large backgrounds 
at lower momenta. Signal electrons are mostly from primary \B decays if they are accompanied by a 
tag electron of opposite charge 
(unlike-sign). Those with a tag of the same charge (like-sign) originate predominantly from  secondary decays of 
charm particles produced in the decay of the other \B meson. Inversion of this charge correlation due to 
\BzBzb\ mixing is treated explicitly, and unlike-sign pairs with both electrons originating from the same 
\B meson are isolated kinematically.
With a small model-dependence on the estimated fraction of primary electrons below $p^*=0.6 \gevc$,
we infer the semileptonic \B branching fraction from the background corrected ratio of unlike-sign 
electron pairs to tag electrons.
 
%-----------------------------------------------------------------------------------------
%-----------------------------------------------------------------------------------------

This measurement is based on data recorded in the year 2000 with the \babar\ detector \cite{BABARDET}
at the \pep2\ energy asymmetric \epem storage ring \cite{PEPII} at SLAC. 
The detector consists of a five-layer silicon vertex tracker (SVT), a 40-layer drift chamber
(DCH), a detector of internally-reflected Cherenkov light (DIRC), and an
electromagnetic calorimeter (EMC)
all embedded in a solenoidal magnetic field of 1.5 T and surrounded by an instrumented flux return (IFR).
To ensure the high quality of the data, we have selected the largest contiguous block of 
events with identical and stable detector conditions in the year 2000, corresponding to an integrated luminosity 
of 4.1\invfb collected at the \FourS\ resonance,
and 0.97\invfb recorded about 40\mev below the \FourS\ peak (off-resonance). 

%-----------------------------------------------------------------------------------------
%-----------------------------------------------------------------------------------------

Multihadron events are selected by requiring a charged track multiplicity
of $N_{\rm ch}> 4$, or $N_{\rm ch}= 4$ plus at least 2 neutral energy deposits above 80\mev 
in the EMC. Track pairs from converted
photons are not included in $N_{\rm ch}$, but count as one neutral particle. 
For further suppression of non-$\BB$ events, we require $R_2<0.6$, where $R_2$ is the ratio of Fox-Wolfram moments $H_2/H_0$ \cite{FoxWolfram}. 

%-----------------------------------------------------------------------------------------
%-----------------------------------------------------------------------------------------

The electron momentum measurement and identification are critical for this analysis. 
For electron candidates we require hits in at least 12 DCH layers,
and a polar angle $\theta$  within the EMC acceptance, i.e. $-0.72 < \cos \theta < 0.92$.
To reduce the contamination from photon conversions and beam-gas background we require the track impact 
parameters in the plane perpendicular to the beams and along the detector axis to be less than 
0.25\cm\ and 3.0\cm, respectively. 

%-----------------------------------------------------------------------------------------
%-----------------------------------------------------------------------------------------

The track finding efficiency $\epsilon_{trk}$ is determined from data as a function of charged multiplicity, 
transverse momentum, polar and azimuthal angle. For signal electrons with $p^*>\pmin \gevc$, the average
efficiency is $(\avgTrkEff \pm 1.1)\%$.

%-----------------------------------------------------------------------------------------
%-----------------------------------------------------------------------------------------

Electron identification is based on the ratio of the energy in the EMC and the track momentum, $E_{\text{EMC}}/p$, 
the shower shape in the EMC, the specific energy loss \dedx in the DCH, and the number of Cherenkov photons 
and the Cherenkov angle measured in the DIRC. 
Muons are eliminated on the basis of \dedx and $E_{\text{EMC}}/p$.
Taking into account the correlations between deposited energy and shape in hadronic 
showers, we combine probability density functions derived from 
data samples for each
discriminating variable to construct the likelihood function $L(\xi)$, \mbox{$\xi \in \{e,\pi,K,p\}$}.
A track is identified as an electron if  
$$
\frac{L(e)}{L(e) + 5\; L(\pi) + L(K) + 0.1\; L(p)} > 0.95 \ .
$$
The weights roughly reflect the relative abundances, their exact values not being 
crucial for electron identification. 

%-----------------------------------------------------------------------------------------
%-----------------------------------------------------------------------------------------

We measure the electron identification efficiency as a function of $p^*$ and 
center-of-mass polar angle $\theta^*$ using radiative Bhabha events. 
For momenta $p^* > 0.6 \gevc$, the average efficiency is 92\% (see Figure 1a).  
However, Monte Carlo simulations indicate that relative to radiative Bhabha events, 
the identification efficiency in \BB\ 
events is reduced between $(4 \pm 2)\%$ at low momenta ($p^* <$ 1\gevc) and $(2 \pm 1)\%$ 
above $p^*=1.6$ \gevc . We correct the measured efficiency for this momentum-dependent difference.

%-----------------------------------------------------------------------------------------
%-----------------------------------------------------------------------------------------

\begin{figure}[ht]
\begin{center}
\includegraphics[height=2.5in]{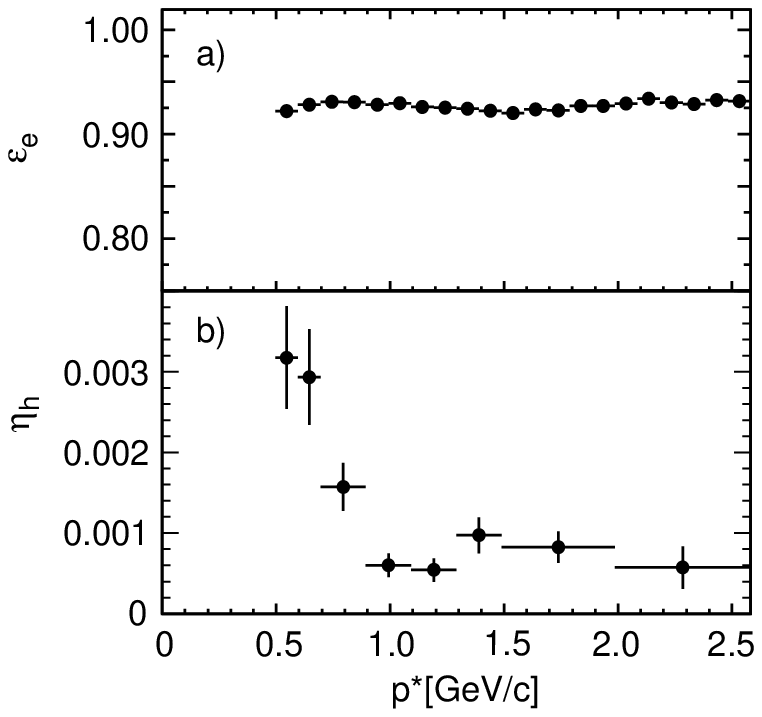} 
\caption {Electron identification efficiency $\epsilon_e$ as obtained from radiative Bhabha events (a) and hadron misidentification rate $\eta_h$ (b) as a function
of $p^*$.}
\label{pid}
\end{center}
\end{figure}

The misidentification rates for pions, kaons, and protons are extracted from control 
samples selected from data.
Figure~\ref{pid}b shows the misidentification probabilities $\eta_h$ per hadronic track,
where the relative abundance of pions, kaons, and protons is taken from \BB\ Monte Carlo simulation. 
The DCH and DIRC contribute significantly at low momenta, while
the performance of the EMC increases with $p^*$. This leads to a minimum of 
0.05\% for $\eta_h$ at $ 1 < p^* < 1.3 \gevc$.
The relative systematic error is estimated to be 15\% from the purities of the 
control samples and the uncertainties in the relative abundances.

%-----------------------------------------------------------------------------------------
%-----------------------------------------------------------------------------------------

The branching fraction analysis makes use of three samples: 
(1) the tag electrons, (2) unlike-sign and (3) like-sign pairs of a tag and a signal electron candidate. 
Misidentified hadrons and electrons from
non-\BB (continuum) events, photon conversions, $\pi^0,\eta \to \gamma \epem$(``Dalitz'') and 
$\jpsi,\psitwos \to \epem$ decays contribute to the background in all three samples. The unlike-sign 
sample also contains pairs of primary and secondary electrons from the same \B meson decay. Further contaminations
to the like- and unlike-sign samples arise from decays of $\tau$ leptons and charmed mesons produced 
in $b \ra c\cbar s$ decays.
Apart from the correction for unlike-sign electron pairs from the same \B, which is performed in bins of $p^*$ only, 
all background corrections are performed in  bins of $p^*$ and polar angle $\theta^*$.

%-----------------------------------------------------------------------------------------
%-----------------------------------------------------------------------------------------

The continuum background is subtracted from all three samples.  It is obtained by normalizing the observed 
off-resonance spectra by the ratio of on- to off-resonance integrated luminosities. The relative systematic 
error in this ratio is estimated to be 0.5\%, attributed to  variations in the detector performance over 
time. The continuum momenta are scaled by the ratio $\sqrt{s_{on}}$/$\sqrt{s_{off}}$ to compensate for the 
0.4\% lower center-of-mass energy.

%-----------------------------------------------------------------------------------------
%-----------------------------------------------------------------------------------------

Electrons from photon conversions and Dalitz decays are identified 
by pairing them with any oppositely charged track with  transverse momentum $p_t>0.1 \gevc$. 
We distinguish the two sources of pairs by the distance $R_{pair}$ of the pair vertex from the detector axis.
Photon conversions are identified by requiring $R_{pair} > 1.6\cm$, a pair invariant mass 
$M_{ee} < 100 \mevcc$, and the transverse and longitudinal distances between the two tracks at the point of 
closest approach $\Delta_{xy} < 0.3 \cm$ and $\Delta_{z} < 1.0 \cm$.
For Dalitz pairs, we require $R_{pair} < 1.6\cm$, $M_{ee} < 200 \mevcc$, $\Delta_{xy}< 0.2 \cm$ and 
$\Delta_{z} < 1.0 \cm$. The momentum- and polar angle-dependent pair finding efficiency, which 
is obtained from a full detector simulation, is low since, in most 
cases, the momentum of the second track is too small to produce a  track in the DCH.  
It varies between 30\% and 40\% for photon conversions and between 20\% and 30\% for Dalitz pairs. 
From a detailed comparison between data and simulation, including the energy 
spectra of the pairs, the relative systematic uncertainties are estimated to be 13\% 
and 19\% for the conversion and Dalitz background rates, respectively. 

%-----------------------------------------------------------------------------------------
% 13
%-----------------------------------------------------------------------------------------

In the unlike-sign sample, electrons from primary and charm decays of the same \B tend
to be produced in opposite directions. Defining $\hat{p}_e^*$ as 
the center of the signal electron momentum bin, this background is reduced by a factor 
of \reductionOaCut\ by imposing the condition 
\begin{equation}
\cos\alpha   > 1.0 - \hat{p}_e^* / \ (\gevc)  \ \ \rm{and} \ \   \cos\alpha> -0.2 \ 
\label{oacut}
\end{equation}
on the opening angle $\alpha$ of \epem pairs, measured in the \FourS\ frame.
Since \B mesons are nearly at rest in this frame, there is no angular correlation between two 
electrons from different $B$ mesons, and the loss in signal efficiency can be calculated on the basis of
geometrical acceptance. 

%-----------------------------------------------------------------------------------------
%-----------------------------------------------------------------------------------------

This selection also eliminates most $\epem$ pairs from inclusive $\B \to \jpsi X$ decays. 
Electron candidates that can be combined with an oppositely charged electron 
to form an invariant mass consistent with the $J/\psi$ hypothesis, 
$2.90 < M_{ee} < 3.15\gevcc$, are excluded from the tag sample if $\cos\alpha<-0.2$.	

%-----------------------------------------------------------------------------------------
%-----------------------------------------------------------------------------------------

\begin{figure}[ht]
\begin{center}
\includegraphics[height=1.72in]{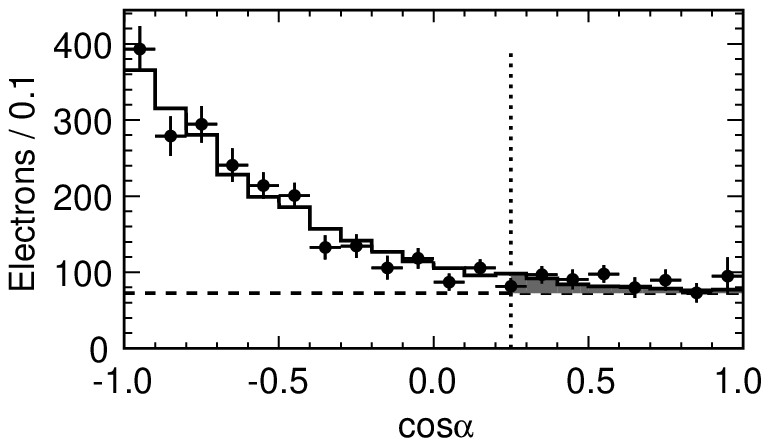} 
\caption {Distribution of the cosine of the opening angle of unlike-sign pairs
for \mbox{$0.7 < p^* < 0.8 \gevc$}. The points represent the data and the histogram is the result of a fit. The shaded area represents the estimated contribution of background electrons, and the vertical dashed line indicates the requirement on the opening angle.}
\label{oafit}
\end{center}
\end{figure}

The contribution of unlike-sign pairs from the same \B decay satisfying Eq.~\ref{oacut} is 
approximately \percCascBg \%.  After subtraction of background contributions from continuum, photon conversions and Dalitz decays,
the observed opening angle distribution (without the requirement) contains a flat contribution from electron pairs from 
different \B mesons and a contribution from electron pairs from the same \B, which peaks at $\cos \alpha = -1$.
The shape of the non-flat background is taken from Monte Carlo simulation and the relative normalization 
of the two contributions is determined by a fit to the data, which is performed separately for each 100 \mevc-wide momentum bin below 1.2 \gevc.
The integral over the fitted non-flat contribution between the minimal allowed value of $\cos \alpha$ and 1 
is taken as the residual 
background (Figure~\ref{oafit}). The very small background above 1.2 \gevc\ (\percCascBgHighP \% of the total contribution) 
is determined from Monte Carlo simulation with a relative uncertainty of 50\%.

%-----------------------------------------------------------------------------------------
%-----------------------------------------------------------------------------------------

We have studied systematic uncertainties in the predicted opening angle distributions  by varying the branching fractions of $\B \to D e \nu$, $\B \to D^* e \nu$, $\B \to D^{**} e \nu$ and non-resonant $\B \to D^{(*)} \pi e \nu$ decays 
by one standard deviation around current average values \cite{PDG00}. 
Based on detailed studies and variations of the fit, the combined systematic error for this background is estimated to be 5\%. 

%-----------------------------------------------------------------------------------------
% 15
%-----------------------------------------------------------------------------------------

\begin{figure}[ht]
\begin{center}
\includegraphics[height=3.4in]{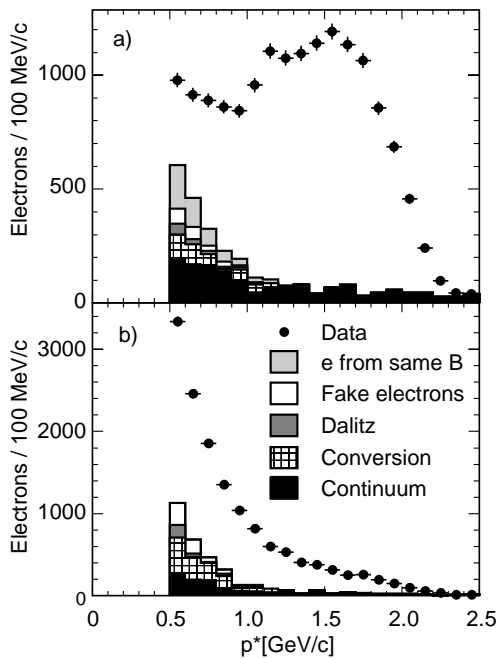} 
\caption{Total measured spectrum (points) and estimated backgrounds (histograms) for signal electron candidates in (a) the 
\uls sample, and (b) the \ls sample.}
\label{rawspectra}
\end{center}
\end{figure}

Figure~\ref{rawspectra} shows the observed momentum spectra and the individual background 
contributions discussed so far, corrected for tracking efficiency; a summary of yields is given in
Table~\ref{grandtable}. Following this initial set of background corrections, the electron yield
is corrected for electron identification efficiency. 

%-----------------------------------------------------------------------------------------
%-----------------------------------------------------------------------------------------
Background contributions from  $\B \to \overline{D} D_{(s)} X$,
\mbox{$D_{(s)} \to e \nu_e Y$} decays and  $\B \to \tau \to e$ decays
are estimated by Monte Carlo simulation, using the currently known branching fractions. 
Combining $\BR(D_s \to X e \nu) = (8.12 \pm 0.68)\%$, which is computed from
the average $D$ branching fraction \mbox{$\BR(D^{0,+} \to X e \nu)$}~\cite{PDG00} and the 
lifetime ratios $\tau_{D^{0,+}} /  \tau_{D_s}$, with 
$\BR (\B \to D_s X ) = (9.8 \pm 3.7) \%$~\cite{cerncombined2001} yields  
$\BR(\B \to D_s \to e) = (0.80 \pm 0.31)\%$.
We take the inclusive branching fraction 
$\BR(\B \to \overline{D} D^{(*)} X)$ to be $(8.2 \pm 1.3)\%$~\cite{cerncombined2001}.
Assuming equal production rates of  $D$ and $ D^{*}$, but allowing for any ratio in the systematic error, 
we arrive at $\BR(\B \to D \to e) = (0.84 \pm 0.21)\%$. 
%------------
To estimate the contribution of electrons from $\tau$ decays,
we use $\BR(\B \to X \tau \nu)$ =  $(2.6\pm 0.2)\%$, 
$\BR(D_s \to \tau \nu)$ =  $(5.79 \pm 2.00) \%$~\cite{dstotau}
and $\BR(\tau \to e \ \nu_e  \overline{\nu}_{\tau}) = (17.83 \pm 0.06)\%$~\cite{PDG00}.
This leads to \mbox{$\BR(\B \to \tau \to e)$} = $(0.565 \pm 0.063) \%$, where the 
$\tau$ lepton originates either directly from a \B decay or a
$\B \to D_s \to \tau$ cascade.
%------------
The background from  $J/\psi$ and $\psi(2S)$ decays into two electrons is also estimated from
Monte Carlo simulation, with $\BR(\B \to J/\psi \to e^+ e^-)$ = $(6.82 \pm 0.38) \times 10^{-4}$
and $\BR(\B \to \psi(2S) \to e^+ e^-)$ = $(3.1 \pm 0.6) \times 10^{-5}$~\cite{PDG00}. 

%-----------------------------------------------------------------------------------------
%-----------------------------------------------------------------------------------------

The tag electron sample is first corrected for continuum background and
hadron misidentification. The remaining background is from secondary decays of charm 
particles and unvetoed $\jpsi \to e^+e^-$ decays. All these contributions are estimated
by Monte Carlo simulation, leading to the background-subtracted number of tag electrons 
$N_{\rm tag}=304{,}048\pm 880\stat\pm 2{,}100\syst$ (Table~\ref{grandtable}),
including a correction for signal loss due to the $\jpsi$-veto.

%-----------------------------------------------------------------------------------------
%-----------------------------------------------------------------------------------------

\begin{table*}[!ht]
\begin{center}
\caption{Electron yield for the three samples and corrections with statistical and systematic errors.}
\begin{tabular}{cccc}
\hline \hline 
 & (1) tag sample & (2) \uls sample , cut on $\alpha$  & (3) \ls sample, all $\alpha$ \\ 
 & $1.4 < p^* < 2.3$ \gevc & $0.6 < p^* < 2.5 \gevc$  & $0.6 < p^* < 2.5 \gevc$  \\ 
\hline
On $\Upsilon(4S)$          &395{,}791$\,\pm\,630$             &  14{,}692$\,\pm\,120  $        & 10{,}838$\,\pm\,110$ \\  
Continuum                  &82{,}073$\,\pm\,590$$\,\pm\,410$  &  1{,}301$\,\pm\,76$$\,\pm\,7$  &939$\,\pm\,64$$\,\pm\,5$ \\  
$\gamma \to e^+ e^- $      & 561$\,\pm\,23$$\,\pm\,140$       &  283$\,\pm\,40$$\,\pm\,37$  &  856$\,\pm\,82$$\,\pm\,110$ \\  
$\eta,\pi^0 \to \gamma e^+e^- $   &   92$\,\pm\,9$$\,\pm\,23$  &  51$\,\pm\,22$$\,\pm\,10$  &  80$\,\pm\,82$$\,\pm\,15$ \\  
Faked $e$   &   1{,}455$\,\pm\,140$$\,\pm\,360$  &  136$\,\pm\,16$$\,\pm\,20$  &  348$\,\pm\,48$$\,\pm\,52$ \\  
$e$ from same B   &     &   317$\,\pm\,7$$\,\pm\,16$  &       \\
\hline 
Yield before and   &   311{,}610$\,\pm\,870$$\,\pm\,570$  &  12{,}603$\,\pm\,150$$\,\pm\,46$  &  8{,}616$\,\pm\,180$$\,\pm\,120$ \\  
after eff. corr.   &     &   14{,}134$\,\pm\,180$$\,\pm\,170$  &  9{,}734$\,\pm\,190$$\,\pm\,200$ \\
\hline   
 $B  \to \tau \to e $   &     &   353$\,\pm\,17$$\,\pm\,42$  &  93$\,\pm\,9$$\,\pm\,11$ \\  
 $B \to D_s  \to e $   &     &   293$\,\pm\,19$$\,\pm\,110$  &  72$\,\pm\,9$$\,\pm\,28$ \\  
 $B \to D  \to e $   &     &   226$\,\pm\,16$$\,\pm\,57$  &  65$\,\pm\,8$$\,\pm\,16$ \\  
 Secondary tags   &  8{,}073$\,\pm\,91$$\,\pm\,2,000$  &  296$\,\pm\,17$$\,\pm\,74$  &  886$\,\pm\,29$$\,\pm\,220$ \\  
 $e$ from $J/\psi$ or $\psi(2S)$   &  1,925$\,\pm\,42$$\,\pm\,120$  &  77$\,\pm\,8$$\,\pm\,5$  &  119$\,\pm\,10$$\,\pm\,7$ \\  
 $e$ removed by $J/\psi$ veto   &   $-$(2{,}435 $\pm$ 50 $\pm$ 220)   &     &    \\
\hline 
 Net $e$ yield   &   304{,}048$\,\pm\,880$$\,\pm\,2{,}100$  &  12{,}890$\,\pm\,180$$\,\pm\,230$  &  8{,}500$\,\pm\,200$$\,\pm\,300$ \\  
\hline \hline 
\end{tabular}
\label{grandtable} 
\end{center}
\end{table*}

\begin{figure}[!h]
\begin{center}
\includegraphics[height=1.8in]{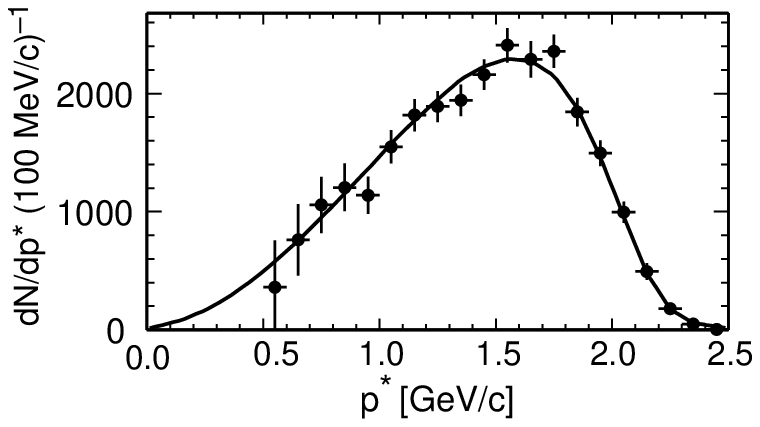} 
\caption {Momentum spectrum of electrons from decays  \mbtox\ after correction 
for efficiencies and external bremsstrahlung, with combined statistical and systematic errors. The curve indicates the fit used for the extrapolation to $p^*=0$.}
\label{fig_spectrum}
\end{center}
\end{figure}

Due to \BzBzb\ flavor oscillations, electrons from primary
\B decays and 
$ \B \to \overline{D} X$, $\overline{D} \to e^-\nu_e Y$ cascades
contribute to both unlike- and like-sign spectra. Denoting the efficiency 
of the opening angle cut as $\epsilon_{\alpha}(p^*)$, their $p^*$ distributions can
be written as 
\begin{equation*}
\begin{aligned}
\frac{1}{\epsilon_{\alpha}(p^*)}\frac{dN^{+-}}{dp^*} &=  
\frac{dN_{B \to X e \nu}}{dp^*} \; (1-\chi) +
\frac{dN_{B \to \overline{D} \to X e \nu}}{dp^*} \: \chi  \ , \\
\frac{dN^{\pm \pm}}{dp^*} & = 
\frac{dN_{B \to X e \nu}}{dp^*} \; \chi +
\frac{dN_{B \to \overline{D} \to X e \nu}}{dp^*} \: (1-\chi)  \ ,
\end{aligned}
\end{equation*}

\noindent where $\chi$ is the product of the \BzBzb\ mixing parameter 
$\chi_0 = 0.174 \pm 0.009$ \cite{PDG00} and $f_0 =\BR(\FourS \to \BzBzb)$.
Since the measured ratio of charged to neutral \FourS\ decays is consistent 
with unity~\cite{CLEOupsilon}, we assume $f_0 = 0.500 \pm 0.025$, 
where the error is taken from~\cite{CLEOupsilon}. We use these linear equations to determine
the primary electron spectrum from \B decays, $dN_{B \to X e \nu}/dp^*$.
Integration of this spectrum between \pmin\ and 2.5 \gevc\ yields $N_{\B \to Xe\nu}=\valNprompt $. 
Using Monte Carlo simulation, we determine the relative efficiency for selecting events with two 
electrons compared to events with a single tag to be $\epsilon_{evt}=(98.0 \pm 0.5)\%$. Together 
with the polar angle acceptance $\epsilon_{geom}=84\%$, we obtain the partial branching fraction 
\begin{equation*}
\begin{aligned}
\BR(\B\to X e \nu,& \;p^*>\pmin\gevc)  = \frac{N_{\B \to Xe\nu}}{N_{tag}\; 
\epsilon_{brem}\;\epsilon_{evt}\;\epsilon_{geom}}  \\
 & = \valBrPartial \ ,
\label{eqbrvis}
\end{aligned}
\end{equation*}
which includes a correction for the small loss of electrons
due to bremsstrahlung in the detector material and the limited momentum resolution,  
$1-\epsilon_{brem} = \valEpsBrem$. The contributions to the systematic error are listed in Table~\ref{tbl_systematics}. 
Figure~\ref{fig_spectrum} shows the momentum spectrum of primary electrons.

%-----------------------------------------------------------------------------------------
%-----------------------------------------------------------------------------------------
To determine the total semileptonic branching fraction, 
we need to extrapolate the spectrum to $p^*=0$.  This is achieved by fitting the 
data to the sum of the spectra from the various exclusive decays.  
We use a parameterization of HQET-derived form factors~\cite{CLEOHQET2,CLEOHQET} to model the decays
$\B \to D e \nu$ and $\B \to D^{*} e \nu$, and the work of Goity and Roberts~\cite{roberts} for non-resonant  
$\B \to D^{(*)} \pi e \nu$ decays. Semileptonic \B decays to $D^{**} e \nu$ and charmless 
mesons are described by the  ISGW2 model~\cite{isgw2}, which is also used as an alternative description for the
processes $\B \to D e \nu$ and $\B \to D^{*} e \nu$. Photon radiation in the final state is modeled by 
PHOTOS~\cite{photos}.
The relative contributions of the different exclusive decay modes are constrained to be within two standard 
deviations of the measured average branching fractions~\cite{PDG00}. 
The best estimate for the extrapolation factor is \mbox{$1 + \kappa = \valKappa$},
where the error accounts for the observed variations of the fit results for
different decay models and branching fractions. This extrapolation leads to a total 
semileptonic branching fraction $\BR_{SL}$ of 
\begin{equation*}
\BRF = \valBr \ .
\end{equation*} 
%-----------------------------------------------------------------------------------------
%-----------------------------------------------------------------------------------------

\begin{table}[!ht]
\caption{Impact of systematic uncertainties on $\BR_{SL}$.}  
\begin{center} 
\begin{tabular}{lr}
\hline \hline
\begin{tabular}{lcc} 
 Source & $\Delta \BR_{SL}  (\%) $ \\ 
\hline 
$e$ efficiency                         & 0.144 \\ 
$\B \to D_s \to e$                     & 0.130 \\
$ \epsilon_{trk}$                      & 0.120 \\ 
extrapolation                          & 0.092 \\ 
$ N_{tag} $                            & 0.075 \\ 
$\B \to D \to e $                      & 0.067 \\ 
mistagged $e$                          & 0.061 \\ 
$f_0 \chi_0                         $  & 0.059 \\ 
$ \epsilon_{evt} $ & 0.054 \\ 

\end{tabular}
&
\begin{tabular}{lcc} 
 Source & $\Delta \BR  (\%) $ \\ 
\hline 
$\B \to  \tau \to e $      & 0.044 \\ 
$\gamma \to e^{+}e^{-}$              & 0.042  \\ 
$ \epsilon_{brem}$                   & 0.039 \\ 
faked $e$                            & 0.024  \\ 
$e$ from same \B                      & 0.022  \\ 
$\pi^0,\eta \to \gamma \; e^{+}e^{-}$& 0.014  \\ 
 continuum                          &  0.008  \\ 
$\jpsi , \psitwos \to e^+e^-$        & 0.003 \\ 
       & \\
\end{tabular}

\\
\hline 
Total  & 0.296\ \ \ \\
\hline \hline 
\end{tabular}
 
\label{tbl_systematics}
\end{center} 
\end{table}

One of the limiting factors of this analysis is the background at low momenta, especially 
semileptonic decays of charmed mesons produced in $b \ra c\cbar s$ decays. As shown in 
Table~\ref{tbl_pmincut}, raising the minimum momentum requirement $p^*_{min}$ reduces the 
systematic uncertainty due to this background substantially, but also increases the error on  
the extrapolation to $p^*=0$. We choose $p^*_{min}=0.6 \gevc$ for the final result,
since the systematic error is comparable with higher values of $p^*_{min}$, while the 
model-dependence is significantly lower.

\begin{table}[!ht] 
\caption{Determination of $\kappa$, $\BR_{SL}$, and the contributions to the systematic error for different signal electron momentum cut-offs. All numbers are stated in percent.} 
\begin{center} 
\begin{tabular}{lcccccc} 
\hline \hline
$p^*_{min}[\gevc] $ & 0.5 & 0.6 & 0.7 & 0.8 & 0.9 & 1.0 \\
\hline
$\kappa$			  & 3.8 & 6.1 & 9.3 & 13.6 & 19.2 & 27.2 \\
$\BR_{SL} $                       & $ 10.79 $  &  $ 10.87$  & $ 10.87 $  & $ 10.82 $  & $ 10.80 $ & $ 10.93 $    \\
$\Delta \BR_{SL}(\gamma,\pi^0) $  & 0.07 & 0.04 & 0.03 & 0.02 & 0.02 & 0.01 \\
$\Delta \BR_{SL}(\epsilon_{trk})$ & 0.12 & 0.12 & 0.12 & 0.12 & 0.12 & 0.12 \\
$\Delta \BR_{SL}(e\; \rm eff.)$     & 0.15 & 0.14 & 0.14 & 0.12 & 0.11 & 0.10 \\ 
$\Delta \BR_{SL}(\B \to D_s )$  & 0.17 & 0.13 & 0.09 & 0.06 & 0.05 & 0.04 \\
$\Delta \BR_{SL}(\B \to D)$  & 0.10 & 0.07 & 0.05 & 0.03 & 0.02 & 0.01 \\
$\Delta \BR_{SL}(\B \to \tau)$  & 0.05 & 0.04 & 0.04 & 0.03 & 0.03 & 0.02 \\
$\Delta \BR_{SL}(\rm extrapolation )$  & 0.06 & 0.09 & 0.13 & 0.19 & 0.25 & 0.33 \\
$\Delta \BR_{SL}(\rm other)$  &        0.15 & 0.14 & 0.14 & 0.15 & 0.15 & 0.17 \\
\hline
$\Delta \BR_{SL}(\rm syst)$  &    0.33 & 0.30 & 0.29 & 0.30 & 0.34 & 0.41 \\
$\Delta \BR_{SL}(\rm stat) $      & 0.21 & 0.18 & 0.16 & 0.16 & 0.15 & 0.15 \\
\hline \hline
\end{tabular} 
\label{tbl_pmincut}
\end{center} 
\end{table}

Based on the work by Hoang~{\it et~al.}~\cite{hoang}, we relate the decay rate and the modulus of the CKM matrix element $V_{cb}$ by 
\begin{equation*}
\begin{aligned}
\Vcb = & (41.9 \pm 2.0) \times 10^{-3}  \\
      & \times \sqrt{\BRFC/ \;0.105}\ \sqrt{1.6 \ps / \tau_{\it B} }. 	
\end{aligned}
\end{equation*}
Using $\tau_B = (1.601 \pm 0.021) \ps$ and 
$\BRFU =(1.7 \pm 0.6)\times 10^{-3}$~\cite{PDG00}  , we obtain \Vcb = \valVcb.

In conclusion, we have used electrons in \FourS\ decays tagged by a high momentum electron to measure \mbox{\BRF = \valBr}.  This measurement is largely model-independent.  The result is in agreement with previous measurements \cite{CLEO96,LEP}, but the systematic uncertainties are reduced.  However, the poorly 
known branching fractions in \B and $D_{(s)}$ decays lead to significant systematic uncertainties in the background subtraction.   
The resulting measurement of \Vcb remains dominated by theoretical uncertainties.
It has recently been shown that non-perturbative effects can be assessed by measurements of moments of inclusive distributions \cite{moments}.

We are grateful for the excellent luminosity and machine conditions
provided by our \pep2\ colleagues, 
and for the substantial dedicated effort from
the computing organizations that support \babar.
The collaborating institutions wish to thank 
SLAC for its support and kind hospitality. 
This work is supported by
DOE
and NSF (USA),
NSERC (Canada),
IHEP (China),
CEA and
CNRS-IN2P3
(France),
BMBF and DFG
(Germany),
INFN (Italy),
NFR (Norway),
MIST (Russia), and
PPARC (United Kingdom). 
Individuals have received support from the 
A.~P.~Sloan Foundation, 
Research Corporation,
and Alexander von Humboldt Foundation.

%\newpage

\end{document}